\documentclass[sn-mathphys-num]{sn-jnl}
\pdfoutput=1

\usepackage{graphicx}%
\usepackage{multirow}%
\usepackage{amsmath,amssymb,amsfonts}%
\usepackage{amsthm}%
\usepackage{mathrsfs}%
\usepackage[title]{appendix}%
\usepackage{xcolor}%
\usepackage{textcomp}%
\usepackage{manyfoot}%
\usepackage{booktabs}%
\usepackage{algorithm}%
\usepackage{algorithmicx}%
\usepackage{algpseudocode}%
\usepackage{listings}%
\usepackage[numbers]{natbib}%
\usepackage{cancel}
\usepackage{slashed}
\usepackage{verbatim}
\usepackage{latexsym}
\usepackage[]{array}
\usepackage{amsxtra}
\usepackage{color}
\usepackage{braket}
\usepackage{units}
\usepackage{ulem} 
\usepackage{hyperref}
\usepackage{textpos}

\newcommand\numberthis{\addtocounter{equation}{1}\tag{\theequation}}
\def\real{{\rm Re}}
\newcommand{\multiN}{\left\lbrace n_\mu \right\rbrace} 
\newcommand{\multiM}{\left\lbrace m_\mu \right\rbrace} 
\newcommand{\indexmu}{\mu} 
\newcommand{\indexnu}{\nu}

\numberwithin{equation}{section}

\theoremstyle{thmstyleone}%

\theoremstyle{thmstyletwo}%

\theoremstyle{thmstylethree}%

\begin{document}

\title[Article Title]{Rotations and boosts of Hermite functions}

\author[1]{\fnm{Maro} \sur{Cvitan}}\email{mcvitan@phy.hr}

\author[2]{\fnm{Predrag} \sur{Dominis Prester}}\email{pprester@math.uniri.hr}

\author[3]{\fnm{Stefano} \sur{Giaccari}}\email{stefano.giaccari@pd.infn.it}

\author*[4]{\fnm{Mateo} \sur{Pauli\v si\'c}}\email{mateo.paulisic@uniri.hr}

\author[5]{\fnm{Ivan} \sur{Vukovi\'c}}\email{vukovic.ac@gmail.com}

\affil[1]{\orgdiv{Department of Physics}, \orgname{Faculty of Science, University of Zagreb}, \orgaddress{\street{Bijeni\v cka cesta 32}, \city{Zagreb}, \postcode{10000}, \country{Croatia}}}

\affil[2]{\orgdiv{Faculty of Mathematics}, \orgname{University of Rijeka}, \orgaddress{\street{Radmile Matej\v ci\'c 2}, \city{Rijeka}, \postcode{51000}, \country{Croatia}}}

\affil[3]{\orgdiv{Dipartimento di Fisica e Astronomia 'Galileo Galilei' e INFN sez. di Padova}, \orgname{Universit\`a di Padova}, \orgaddress{\street{Via Marzolo 8}, \city{Padova}, \postcode{35131}, \country{Italy}}}

\affil[4]{\orgdiv{Faculty of Physics}, \orgname{University of Rijeka}, \orgaddress{\street{Radmile Matej\v ci\'c 2}, \city{Rijeka}, \postcode{51000}, \country{Croatia}}}

\affil[5]{\city{Vienna}, \country{Austria}}

\abstract{We provide transformation matrices for arbitrary Lorentz transformations of multidimensional Hermite functions in any dimension. These serve as a valuable tool for analyzing spacetime properties of MHS fields, and aid in the description of the relativistic harmonic oscillator and digital image manipulation. We also focus on finite boosts and rotations around specific axes, enabling us to identify the Lorentz Lie algebra generators. As an application and to establish a contact with the literature we construct a basis in which the two dimensional rotation operator is diagonal. We comment on the use of hypergeometric functions, the Wigner d-functions, Kravchuk polynomials, Jacobi polynomials and generalized associated Legendre functions.}


\maketitle
\begin{textblock*}{5cm}(10.5cm,-18cm) 
	Preprint number: ZTF-EP-24-05
\end{textblock*}

\section{Introduction}
Modern theoretical high-energy physics is based on the assumption that on flat Minkowski background elementary fields and particles are classified by irreducible unitary representations of the Poincar\'e group or its generalizations, such as the conformal group or the super-Poincar\'e one. In the seminal papers \cite{Wigner:1939cj, WignerDeutch, Bargmann:1948ck}  the irreducible unitary representations of the $4d$ Poincar\'e group were classified into massive, massless and tachyonic ones, apart from the zero-momentum unfaithful representations. The massless representations are in turn divided into discrete/finite spin ones, characterized by helicity taking a finite set of values so that they have a finite number of components, and  continuous/infinite  spin ones, (see e.g. \cite{Bekaert:2006py} for a modern treatment of the subject), characterized by a  continuous value of the second quartic Casimir. Whereas traditionally the latter representations, with their infinite number of degrees of freedom at any point in spacetime, have been neglected in field theory, in recent years they have been the object of novel interest \cite{Schuster:2013pxj, Schuster:2013vpr, Schuster:2013pta, Schuster:2014hca, Rivelles:2016rwo, Bekaert:2017khg, Alkalaev:2017hvj, Buchbinder:2020nxn, Schuster:2023xqa, Schuster:2023jgc} due to their enticing properties which seem to fit well into such distinctively stringy regimes as the tensionless limit, where the onset of higher spin gravity is expected. In fact, the spectrum of helicities of a continuous-spin particle coincides with the one in higher-spin gravity and the presence of a dimensionful continuous parameter might help circumvent the no-go theorems for interacting higher spin particles in flat background in a way similar to what happens in the presence of a cosmological constant.

On the other hand, there are situations where it is not completely clear how the physical solutions can retain Lorentz covariance without breaking unitarity. One celebrated such case is the relativistic harmonic oscillator, considered as a model for bound states of quarks in a relativistic setting \cite{Feynman:1971wr, Kim:1973dc} (see also \cite{Kim:2020eku} for a review), but whose eigenvalue equation plays also a crucial role in determining the spectrum of string theory. It was pointed out \cite{Bars:2008wb} that the standard approach in string theory boils down to considering solutions of the relativistic harmonic oscillator equation in a spacelike sector of every normal mode of the relativistic string. On the one hand this ensures the elimination of most tachyonic modes and the absence of negative norm states upon application of covariant quantization, on the other hand it is not to be excluded that other solutions may exist. In particular it was pointed out there are solutions supported on mixed spacetime and timelike sectors that are completely free of negative norm states and realize infinite dimensional unitary representations of the Lorentz group. Such representations were first considered by Dirac \cite{Dirac1945} in 1944 on a suitably defined space of infinite sums of polynomials of a real variable with the coefficients in the sums named ``expansors''. In this paper we are interested in an infinite dimensional unitary representations of the Lorentz group constructed using fields expanded in Hermite functions.

The need for studying such infinite dimensional unitary representations of the Lorentz group comes in the context the Moyal-Higher-Spin (MHS) formalism, defined and developed in \cite{Bonora:2018uwx,Bonora:2018eot,Cvitan:2021yvf,Cvitan:2021qvm}. It is an approach to study higher spin theory, built on a $2d$-dimensional manifold $\mathcal{M}\times\mathcal{U}$ with coordinates $(x^a,u_b)$, where $\mathcal{M}$ represents Minkowski spacetime, and $\mathcal{U}$ an auxiliary space of equal dimension. Fields $\Phi(x,u)$ that live on such a manifold depend on both $x$ and $u$ coordinates.

To better understand the purely spacetime content of the MHS fields, one needs a way of integrating out the dependency on the auxiliary coordinates, while ensuring that physical observables remain well defined. This motivates expanding the MHS fields in an orthonormal basis of functions in the auxiliary space, which ensures that functionals quadratic in the field variables are finite;

\begin{equation}
\Phi(x,u) = \sum_{M} \phi^M(x) f_M(u)\,,
\end{equation}
where $\{f_M(u)\}$ are orthonormal functions on the auxiliary space. A particularly convenient choice for the basis are Hermite functions defined below. The MHS models developed primarily in \cite{Cvitan:2021yvf} incorporate the Poincar\'e group of symmetries of spacetime, of which the Lorentz group is a subgroup, thus creating a necessity to understand the behavior of basis functions under Lorentz transformations $f_M'(u) = f_M(\Lambda^{-1} u)$, particularly in the infinitesimal case.

The defining representation of the Lorentz group are finite dimensional matrices $\Lambda^\mu{}_\nu$, which are not unitary, instead satisfying\footnote{Einstein's summation convention is understood and we use the $\eta \sim (-,+,...,+)$ signature.}
\begin{equation}\label{LorentzDef}
\Lambda^\mu{}_\alpha\,\eta_{\mu\nu}\,\Lambda^\nu{}_\beta = \eta_{\alpha\beta}.
\end{equation}
In this work, we build a representation of the Lorentz group on a space of multi-dimensional Hermite functions, which is unitary by construction, albeit, infinite-dimensional. Since the group of rotations is a subgroup of the Lorentz group, the results we obtain can also be used to represent Euclidean rotations on the space of Hermite functions in an arbitrary number of dimensions. Results covering the case of one-dimensional boosts were obtained in \cite{Ruiz:1974uf, Rotbart:1981xf} in the context of the relativistic harmonic oscillator, while the problem of Euclidean rotations of Hermite functions was also studied in the context of computer graphics \cite{Park2008, ParkAccurate2009, Reynolds2018}. The complete solution for an arbitrary Lorentz transformation was so far not available.

We start with a reminder on representing groups on spaces of functions and outline the method we used to construct the representation. Afterwards, we specialize to Hermite functions and the Lorentz group and provide explicit representation matrices for an arbitrary Lorentz transformation, as well as for particular cases of boosts in specific directions and rotations around specific axes in $D=4$. From the finite case, we find the generators of the Lorentz Lie algebra in $D=4$, calculate the Casimir elements, and as an application we discuss a basis for the vector space of Hermite functions for which the rotation operator around the $z$-axis is diagonal. The appendices contain details necessary for the diagonalization procedure including the derivation of the result with the help of the Kravchuk polynomials and a presentation of the result in terms of the Wigner $d$-functions. Relation to hypergeometric functions, Jacobi polynomials and generalized associated Legendre functions is also exhibited.

\section{Representing a group on a function space}
Let a group $G$ be represented on $\mathbb{R}^D$ by linear operators $g$. The action on functions on $\mathbb{R}^D$ is given by
\begin{equation}
h'(x) \equiv h(g^{-1}x)\,.
\end{equation}
Let us choose a complete orthonormal real basis of functions $f_N (x)$ which will span $L^2(\mathbb{R}^D)$, indexed by a (possibly composite) index $N$. The orthonormality and completeness conditions are given by
\begin{align}
\int d^Dx\, f_N(x)f_M(x) = \delta_{NM},\qquad
\sum_{N}f_N(x)f_N(y) = \delta^{(D)}(x-y)\label{orthogonality}\,.
\end{align}
We can expand an arbitrary element $h(x)\in L^2(\mathbb{R}^D)$ in the chosen basis as
\begin{equation}
h(x) = \sum_N h^N f_N (x)\,.
\end{equation} 
Since a transformed function $h'(x) = h(g^{-1}x)$ can be expanded in two ways
\begin{align}
h'(x) =& \sum_N h^N f_N (g^{-1}x)=\sum_M h'^M f_M(x)\,,\label{function_expansion}
\end{align}
the orthogonality of the basis functions (\ref{orthogonality}) can be used to express
\begin{align}
h'^{M} = \sum_{N} h^N \left(\int d^Dx\, f_M (x) f_N (g^{-1}x)\right)\,,
\end{align}
which furnishes a representation of $G$ denoted by\footnote{The constructed matrices $ D_{N}^M (g) $ are real since the choice of the basis functions $f_N(x)$ was real. The procedure can be generalized to complex valued functions.} 
\begin{align}\label{defD}
D^M_N(g) \equiv \int d^Dx\, f_M (x) f_N (g^{-1}x)\,.
\end{align}
The transformation between function coefficients can now be written as
\begin{align}
\label{component_transform}
h'^{M} = \sum_{N} D_{N}^M (g) h^N
\end{align}
We indeed have a preserved group structure, since for $h''(x) = h(g_1^{-1}g_2^{-1} x) = h(g_3^{-1}x)$ one can obtain
\begin{equation}
D^N_{K}(g_3)=  \sum_{M}D_M^{N}(g_2)  D^M_{K}(g_1)\,.
\end{equation}
In case of special pseudo-orthogonal groups $SO(p,q)$ (e.g. the Lorentz group or the rotation group), the obtained matrices $D^M_N(g)$ will be unitary. To prove unitarity (i.e.\ orthogonality in our case) we notice an equality
\begin{align}\label{changed_coordinates}
&\int d^Dx \, f_N(g^{-1}x) f_M(g^{-1}x) = \int d^Dy \, f_N(y) f_M(y) = \delta_{NM}\,,
\end{align}
owing to the Jacobian being equal to unity. By using equation (\ref{defD})
\begin{equation}
\label{basis_transform}
f_N(g^{-1}x) = \sum_M D^M_N(g) f_M(x)\,,
\end{equation}
we obtain
\begin{equation}
\sum_J D^J_N D^J_M = \sum_J (D^T)^N_J D^J_M = \delta_{NM}\,.
\end{equation}
Below we will use this approach to construct a unitary representation of the Lorentz group.
\section{Representation of the Lorentz group on $L^2(\mathbb{R}^{D})$}
\subsection{Hermite functions and the generating integral}
Our choice for the basis of $L^2(\mathbb{R}^{D})$ are multi-dimensional Hermite functions defined below. Partial results for the representation matrices of the Lorentz group on Hermite functions were obtained by Ruiz \cite{Ruiz:1974uf} and generalized by Rotbart \cite{Rotbart:1981xf} in the context of finding transformed eigenfunctions of a relativistic quantum harmonic oscillator. Their results cover the case of one-dimensional boosts. 
The main idea we take from their calculations is to integrate a product of generating functions and find the sought-for result in the subsequent expansion. Our approach is more general and enables us to calculate the representation matrices for arbitrary elements of the Lorentz group. Here we note that the results we obtain cannot be simply deduced from the results \cite{Ruiz:1974uf} and \cite{Rotbart:1981xf} e.g.\ using tricks involving covariance. The reason for this can be traced to the form of the weight function. Namely, if for simplicity we restrict for the moment to the 
$4$-dimensional case, our basis functions are weighted using $\exp( -t^2 -x^2 -y^2 -z^2 )$ which makes the basis functions localized, but on the other hand complicates their transformation properties since $-t^2 -x^2 -y^2 -z^2$ is not invariant under Lorentz transformations (cf.\ footnote \ref{footnotexxnx}). Hence, here we do the calculation assuming a general form of a Lorentz transformation matrix from the start.

To begin, we introduce $H_{n}(x)$, the (physicists') Hermite polynomials
\begin{equation}
H_{n}(x)=(-1)^{n} e^{x^{2}} \frac{d^{n}}{d x^{n}} e^{-x^{2}}
\,,
\end{equation}
where the index $n$ can attain arbitrary non-negative integer values.  Next, the Hermite functions are defined as the Hermite polynomials multiplied by the suitable normalization (and weight) factor as
\begin{equation}
\label{Hermite_functions}
f_{n}(x) = \frac{1}{c_n} e^{-\frac{x^2}{2}} H_{n}(x)\,.
\end{equation}
where
\begin{equation}
c_n = \sqrt{2^n n! \sqrt{\pi}}
\end{equation}
Importantly, they are orthonormal and complete on $\mathbb{R}$
\begin{equation}
\int_{-\infty}^{\infty} dx\,f_n(x)f_m(x) = \delta_{nm}\,,\,\quad \label{completeness_Hermite}
\sum_{n=0}^\infty f_n(x)f_n(y) = \delta(x-y)\,.
\end{equation}
We define a multi-dimensional Hermite function as a product\footnote{An alternative generalization to higher dimensions is provided by Grad \cite{Grad:1949}, which we have used in \cite{Cvitan:2021qvm} in the Euclidean approach to MHS theory. However, while Grad's generalization provides a covariant way to represent Euclidean rotations, it is not particularly suitable for representing the Lorentz group, since their basis functions mix rank under Lorentz boosts.}
\begin{equation}
\label{multi_dimensional_hermite}
f_{n_0\dots n_{D-1}}(x) \equiv f_{n_0}(t)f_{n_1}(x^1)\cdots f_{n_{D-1}}(x^{D-1}) \equiv f_{\multiN}(x)\,,
\end{equation}
with a multi-index notation, $\multiN = \{n_0\dots n_{D-1}\}$. 
To calculate the representation matrices for $\Lambda \in \text{SO}(1,D-1)$ we need to use (\ref{defD}).
In order to evaluate the above integral, we introduce the generating functions for Hermite polynomials
\begin{equation}
e^{2xq - q^2} = \sum_{m=0}^\infty H_{m}(x) \frac{q^{m}}{m!}\,,
\end{equation}
and Hermite functions
\begin{equation}
\label{generating_hermite}
E_1(x,q)\equiv e^{2xq - q^2-x^2/2} = \sum_{m=0}^\infty c_{m}\frac{q^{m}}{m!}f_{m}(x)
\end{equation}
which are obtained by multiplying generating functions for Hermite polynomials by $e^{-x^2/2}$. We can easily generalize to $D$ dimensions
\begin{equation}
\label{generating_function_4dim}
E(x,q) \equiv E_{1}(x^0,q^0)E_{1}(x^1,q^1)\ldots E_{1}(x^{D-1},q^{D-1})\,.
\end{equation}
Next, we multiply two generating functions (\ref{generating_function_4dim}), integrate the product,
\begin{align}
\label{master_integral0}
I(p,q,\Lambda) =
\int d^{D}x\,E(x,q)E(\Lambda^{-1}x,p)
\end{align}
and expand it as a series in the generating variables
\begin{align}
\label{master_integral}
I(p,q,\Lambda) =&
\sum_{\multiM} \sum_{\multiN} 
\left[ \prod_{{\nu}} c_{m_\nu} c_{n_\nu}  
	\frac{(q^{\nu})^{m_{\nu}}}{m_{\nu}!} 
	\frac{(p^{\nu})^{n_{\nu}}}{n_{\nu}!} \right]
D^{\multiM}_{\multiN}(\Lambda)\,.
\end{align}	
We find on the right hand side of (\ref{master_integral}) the transformation matrices i.e.\ the coefficients $D^{\multiM}_{\multiN}(\Lambda)$ defined by (\ref{defD}) i.e.\ 
\begin{align}
\label{Dmatrix1}
D^{\multiM}_{\multiN}(\Lambda) = \int d^{D}x\, f_{\multiM}(x)f_{\multiN}(\Lambda^{-1}x)
\end{align}
Below, we obtain the integral on the left hand side of (\ref{master_integral}) for any Lorentz transformation. \footnote{In \cite{Ruiz:1974uf,Rotbart:1981xf} light-cone coordinates were employed and  this integral was evaluated for the case of one-dimensional boosts.}
\subsection{Generating integral for an arbitrary Lorentz transformation}
To solve (\ref{master_integral0}) we will rewrite it to recognize the form of a Gaussian integral. In addition to $q^\mu \equiv (q^0, {q^i})$ and $p^\mu \equiv (p^0, {p^i})$,	we introduce auxiliary variables
\begin{equation}
{u^\mu \equiv (1,0,\ldots,0)}\,,
\end{equation}
and define
\begin{equation}
\eta^\textrm{eucl}_{\mu\nu} = \eta_{\mu\nu}+2u_{\mu} u_{\nu}\,, 
\end{equation}
i.e.\ $\eta \sim (-+++)$ and $\eta^\textrm{eucl} \sim (++++)$. 
Even though $q^\mu$, $p^\mu$, and $u^\mu$ resemble components of a Lorentz vector, they do not change under Lorentz transformations (in other words, we never apply Lorentz transformations to them since they are tied to the laboratory frame in which we work). Their purpose is to be a placeholder and enable a more concise notation.\footnote{\label{footnotexxnx}Eg. $t^2 + x^2 + y^2 + z^2 = x_\mu x^\mu + 2(u_\mu x^\mu)^2$} The generating function (\ref{generating_function_4dim}) now becomes
\begin{equation}
E(x,q) = \exp\left[ 2x^\mu q^\nu \eta^\textrm{eucl}_{\mu\nu}  - q^\mu q^\nu \eta^\textrm{eucl}_{\mu\nu} -
\frac12 x^\mu x^\nu \eta^\textrm{eucl}_{\mu\nu}  \right]\,.
\end{equation}
We rewrite the integral (\ref{master_integral0}) as
\begin{equation}
\int d^{D}x\,E(x,q)E(\Lambda^{-1}x,p)=\int d^{D}x\, \exp[-\frac{1}{2}x^\alpha A_{\alpha\beta}x^\beta + J_\alpha x^\alpha + C]\,,
\end{equation}
with
\begin{align}
&A_{\alpha\beta} 
=\, 2 \left( \eta_{\alpha\beta}+u_\alpha u_\beta\right) + 2(\Lambda^{-1})_{0\alpha}(\Lambda^{-1})_{0\beta } \\
&J_\alpha =\, 2 p^\mu \eta^\textrm{eucl}_{\mu\nu}(\Lambda^{-1})^\nu{}_{\alpha} + 2q^\nu\eta^\textrm{eucl}_{\nu\alpha}\\
&C = - q^\mu q^\nu \eta^\textrm{eucl}_{\mu\nu} - p^\mu p^\nu \eta^\textrm{eucl}_{\mu\nu}\,.
\end{align}
The result of the integral can now easily be obtained since it is of the Gaussian form  \footnote{\label{footnote_gaussian_integral}See e.g.\ ch.\ 9 in \cite{weinberg1995quantum}
	\begin{equation}
	\int d^{D}x \exp\left[-\frac{1}{2}x^T A x + J^T x + C\right] = \sqrt{\frac{(2\pi)^{D}}{\det A}}\exp\left[\frac{1}{2}J^T A^{-1} J + C\right]
	\end{equation}}
\begin{align}
\label{master_integral_result1}
I(p,q,\Lambda) 
=&	
\frac{(2\pi)^{{{D/2}}}}{\sqrt{\det A}}\exp\left[\frac{1}{2} J_\alpha [A^{-1}]^{\alpha\beta}J_\beta + C\right]\,.
\end{align}
We find the determinant $\det A = 2^{D}\,\left( (\Lambda^{-1})^{00} \right)^2$, and the inverse
\begin{align}
\label{inversA}
[A^{-1}]^{\alpha\beta} 
=& \frac{\eta^{\alpha\beta}}{2} + 
\frac{
	(\Lambda^{-1})^{0\alpha}{u}^{\beta} +  
	(\Lambda^{-1})^{0\beta}{u}^{\alpha}
}{2(\Lambda^{-1})^{00}}\,.
\end{align}
The generating integral (\ref{master_integral_result1}) is now given by
	\begin{equation}
	I(p,q,\Lambda)=\frac{\pi^{D/2}}{\left| (\Lambda^{-1})^{00} \right| }\exp\left[
	2 p^0 q^0 M_{00} +
	2 p^0 p^i M_{i0} +
	2 q^0 q^j M_{0j} +
	2 p^i q^j M_{ij}
	\right]\label{Master_integral2}
	\end{equation}
	with the matrix $M$ defined by components
	\begin{align}\label{Master_integral2M}
	M_{00} &= -\frac{1}{(\Lambda^{-1})^{00}}\,, \qquad M_{i0} = -\frac{(\Lambda^{-1})_{i0}}{(\Lambda^{-1})^{00}}\\\nonumber
	M_{0j} &= -\frac{(\Lambda^{-1})_{0j}}{(\Lambda^{-1})^{00}}\,,\qquad M_{ij} = (\Lambda^{-1})_{ij}-\frac{(\Lambda^{-1})_{i0}(\Lambda^{-1}){}_{0j}}{(\Lambda^{-1})^{00}}\,.
	\end{align}
The formulas (\ref{Master_integral2})-(\ref{Master_integral2M}) represent the first important result of this paper. It furnishes, through an expansion performed in Section \ref{sec:extracting} below, a transformation matrix for Hermite functions for an arbitrary Lorentz transformation. 

As a consistency check we calculate $I(p,q,\Lambda)$ for the case of one-dimensional boosts, eg. $v^x=v\neq0$ while $v^y=v^z=0$. That case was covered by \cite{Rotbart:1981xf} in their formula (10). To get their result we can set
\begin{equation}
p^\mu = (p^0, p^1),\quad q^\mu = (q^0, q^1)\,.
\end{equation}
The resulting integral
\begin{align}
I(p,q,v) = \pi^2\sqrt{1-v^2}\exp\left[-2p^0 p^1 v + 2q^0 q^1 v + 2p^0 q^0 \sqrt{1-v^2} + 2p^1q^1 \sqrt{1-v^2}\right]
\end{align}
agrees with \cite{Rotbart:1981xf} after appropriate identifications.

\subsection{Extracting the representation matrices}\label{sec:extracting}
From the expression (\ref{master_integral}) we can see that to extract the representation matrices, we will have to expand $I(q,p,\Lambda)$ in powers of $p$ and $q$.
\footnote{For convenience, we report the formula (\ref{master_integral_expansion}) in $D=4$ with all the indices spelled out \begin{align*}
	I(p,q,\Lambda) =& 
	\sum_{n_0=0}^\infty \cdots 
	\sum_{n_{3}=0}^\infty
	\sum_{m_0=0}^\infty \cdots 
	\sum_{m_{3}=0}^\infty	
	\\
	&\frac{(p^0)^{n_0}}{n_0!}\frac{(p^1)^{n_1}}{n_1!}\frac{(p^2)^{n_2}}{n_2!}\frac{(p^3)^{n_3}}{n_3!}\frac{(q^0)^{m_0}}{m_0!}\frac{(q^1)^{m_1}}{m_1!}\frac{(q^2)^{m_2}}{m_2!}\frac{(q^3)^{m_3}}{m_3!}\times\\
	&\times\left[\frac{\partial^{n_0 + n_1 + n_2 + n_3 + m_0 + m_1 + m_2 + m_3}}{(\partial p^0)^{n_0} (\partial p^1)^{n_1} (\partial p^2)^{n_2}(\partial p^3)^{n_3}(\partial q^0)^{m_0} (\partial q^1)^{m_1} (\partial q^2)^{m_2} (\partial q^3)^{m_3}}I(p,q,\Lambda)\right]\Bigg|_{p=q=0}
	\end{align*}}
\begin{equation}\label{master_integral_expansion}
I(p,q,\Lambda) = \sum_{\multiN} \sum_{\multiM}  
\left[ \prod_{\mu}  
	\frac{(p^{\mu})^{n_{\mu}}}{n_{\mu}!}
	\frac{(q^{\mu})^{m_{\mu}}}{m_{\mu}!}
\right] 
\left[
\frac{\partial^{\sum n_{\mu}+\sum m_{\mu}}
	I(p,q,\Lambda)}{
	\prod_{\mu} 
	(\partial p^{\mu})^{n_{\mu}}
	(\partial q^{\mu})^{m_{\mu}}
}\right]\Bigg|_{p=q=0}
\end{equation}
The representation matrices are then given by
\begin{align}
D^{\multiM}_{\multiN}(\Lambda)=&\frac{1}{\prod_{\mu}  c_{m_{\mu}} c_{n_{\mu}}   }
\left[\frac{\partial^{\sum n_{\mu}+\sum m_{\mu}}
	I(p,q,\Lambda)}{
	\prod_{\mu} 
	(\partial p^{\mu})^{n_{\mu}}
	(\partial q^{\mu})^{m_{\mu}}	
}\right]
\Bigg|_{p=q=0}
\label{Dmatrix2}
\end{align}
We expand the exponent in (\ref{Master_integral2}) as
\begin{equation}
I(p,q,\Lambda) = \frac{\pi^{D/2}}{\left| (\Lambda^{-1})^{00} \right| }
e^{f(p,q,\Lambda)} = \frac{\pi^{D/2}}{\left| (\Lambda^{-1})^{00} \right| } \sum_r \frac{1}{r!}f(p,q,\Lambda )^r\label{ExpansionMethod}\,.
\end{equation}
Since  in (\ref{Dmatrix2}) we put $p$'s and $q$'s to zero at the end of the calculation, in order to have a nonzero contribution, the numbers of derivatives in (\ref{Dmatrix2}) must be related to the powers $f(p,q,\Lambda)^r$ in the following way
\begin{align}\label{termr}
r=m_0+\sum_i n_i = n_0 + \sum_i m_i
\end{align}
This implies that there is only one term in the sum over $r$ in (\ref{ExpansionMethod}) that contributes in (\ref{Dmatrix2}). Therefore, to get $D^{\multiM}_{\multiN}(\Lambda)$ we need to write down an expression for $f(p,q,\Lambda)^r$, identify the term containing $\prod_{\mu}  (p^{\mu})^{n_{\mu}} (q^{\mu})^{m_{\mu}}$ and divide by $\prod_{\mu}  c_{m_{\mu}} c_{n_{\mu}}$. We obtain
\begin{align}\label{DmatrixGenD}
D_{\multiN}^{\multiM}(\Lambda) = &\,
\frac{1}{\left| (\Lambda^{-1})^{00} \right| }
\delta_{-n_0+\sum_{i} n_{i}}^{-m_0+\sum_{i} m_{i}}
\sqrt{\prod_{\mu}  n_{\mu}! m_{\mu}!} 
{\sum_{\left\lbrace n_{\mu\nu} \right\rbrace}}'
\prod_{\mu\nu} \frac{ \left(M_{\mu\nu} \right)^{n_{\mu\nu}} }{n_{\mu\nu}!} 
\end{align}
The symbol ${\sum'_{\left\lbrace n_{\mu\nu} \right\rbrace}}$ denotes $D \times D$ sums over each $n_{\mu\nu}$ from 0 to $\infty$, with the prime $'$ on the sum denoting that the values of $n_{\mu\nu}$ are in addition constrained in the following way:
\begin{align}\label{DmatrixGenDconstr}
m_{0} = \sum_{\nu} n_{0\nu} \, , \quad
n_{i} = \sum_{\nu} n_{i\nu}  \\\nonumber
n_{0} = \sum_{\mu} n_{\mu{}0} \, , \quad
m_{j} = \sum_{\mu} n_{\mu{}j} \,.
\end{align}
Here, we note that these constraints would look more symmetric if $n_0$ and $m_0$ were interchanged. This ``interchange'' is visible also in (\ref{termr}) and in the fact that there are minuses in front of $n_0$ and $m_0$ in the Kronecker symbol in (\ref{DmatrixGenD}).
We trace this to the exponent (\ref{Master_integral2}) where the terms $p_0 p_i$ and $q_0 q_i$ appear. The reason for that is the pseudo-orthogonality of the Lorentz group; such terms are not present in case of euclidean rotations as we show in the subsection below. It is also useful to note that (\ref{termr}) and (\ref{DmatrixGenDconstr}) imply 
\begin{align}
\sum_{\indexmu} n_{\indexmu} + \sum_{\indexmu} m_{\indexmu} &= 2r \\\nonumber
\sum_{\indexmu \indexnu} n_{\indexmu{}\indexnu} &= r
\end{align}

\subsection{Representation matrices for boosts in $D=4$}
For a boost in a single direction (e.g.\ $x$) with velocity $v$, we find the same result as in \cite{Rotbart:1981xf} but generalized to 4 dimensions
\begin{equation}
	I(p,q,v) = \frac{\pi^2}{\gamma}\exp\left[2(q^0 q^1 -p^0 p^1)v + 2(p^0 q^0+ p^1q^1)\sqrt{1-v^2}+ 2p^2q^2 + 2p^3q^3\right]\,.
\end{equation}
By expanding and identifying the coefficients (or just using (\ref{DmatrixGenD})), we find that the transformation matrices for boosts in the $x$ direction are given by\footnote{Here, we used the following notation and conventions for boosts. E.g.\ for boosts in the $z$ direction:
		\begin{align}
		D_{n_0n_1n_2n_3}^{m_0m_1m_3m_3}(v\hat{z}) \equiv D_{n_0n_1n_2n_3}^{m_0m_1m_3m_3}(\Lambda(v\hat{z}))
		\end{align}
		with
		\begin{align}
		\Lambda(v\hat{z}){}^\mu{}{}_\nu \equiv 
		\begin{pmatrix}
		\gamma & 0 & 0 & v\gamma \\
		0 & 1 & 0 & 0 \\
		0 & 0 & 1 & 0 \\
		v\gamma & 0 & 0 & \gamma
		\end{pmatrix}.
		\end{align}}
\begin{align}
\label{large_boost_x}
&D_{n_0n_1n_2n_3}^{m_0m_1m_3m_3}(v\hat{x}) = \,\,\delta_{-n_0+n_1+n_2+n_3}^{-m_0+m_1+m_2+m_3}\delta^{m_2}_{n_2}\delta^{m_3}_{n_3}\sqrt{\frac{m_1!n_0!}{n_1!m_0!}}\,\times\nonumber\\ &\sum_{j=0}^{m_0}\binom{m_0}{j}\binom{n_1}{m_1-j}(-1)^{n_1-m_1+j}\sqrt{1-v^2}^{m_1+m_0+1-2j}v^{2j-m_1+n_1}\,.
\end{align}
Particular cases of boosts in $y$ and $z$ directions lead to
\begin{align}
\label{large_boost_y}
&D_{n_0n_1n_2n_3}^{m_0m_1m_3m_3}(v\hat{y}) = \,\,\delta_{-n_0+n_1+n_2+n_3}^{-m_0+m_1+m_2+m_3}\delta^{m_1}_{n_1}\delta^{m_3}_{n_3}\sqrt{\frac{m_2!n_0!}{n_2!m_0!}}\times\nonumber\\ &\sum_{j=0}^{m_0}\binom{m_0}{j}\binom{n_2}{m_2-j}(-1)^{n_2-m_2+j}\sqrt{1-v^2}^{m_2+m_0+1-2j}v^{2j-m_2+n_2}
\end{align}
\begin{align}
\label{large_boost_z}
&D_{n_0n_1n_2n_3}^{m_0m_1m_3m_3}(v\hat{z}) = \,\,\delta_{-n_0+n_1+n_2+n_3}^{-m_0+m_1+m_2+m_3}\delta^{m_1}_{n_1}\delta^{m_2}_{n_2}\sqrt{\frac{m_3!n_0!}{n_3!m_0!}}\times\nonumber\\ &\sum_{j=0}^{m_0}\binom{m_0}{j}\binom{n_3}{m_3-j}(-1)^{n_3-m_3+j}\sqrt{1-v^2}^{m_3+m_0+1-2j}v^{2j-m_3+n_3}\,.
\end{align} 

The matrices (\ref{large_boost_x})-(\ref{large_boost_z}) are unitary by construction, and they are infinite dimensional as each index $n,m$ ranges from $0$ to $\infty$. The number $N=-n_0+n_1+n_2+n_3=-m_0+m_1+m_2+m_3$ is invariant (in the sense that $SO(1,3)$-transformed basis functions are linear combinations of basis functions of the same $N$), and can be used to reduce the matrices (and basis functions) into sectors labeled by $N$. It is, however, easy to see that each such sector is in itself infinite dimensional; there is an infinite number of ways to combine one negative and three positive integers to obtain the same $N$. For example, consider a function $f_{0000}(x)$, and boost it in the $x$ direction. We can apply the transformation rule (\ref{basis_transform})
\begin{align}
f'_{0000}(x) = f_{0000}(\Lambda^{-1}x) =& \sum_{m_0=0}^\infty\sum_{m_1=0}^\infty\sum_{m_2=0}^\infty\sum_{m_3=0}^\infty D_{0\,0\,0\,0}^{m_0m_1m_2m_3}(v \hat{x})f_{m_0m_1m_2m_3}(x)\nonumber\\
=&\sqrt{1-v^2}\sum_{m_0=0}^\infty\sum_{m_1=0}^\infty \delta_{m_0}^{m_1}v^{m_1}f_{m_0m_100}\nonumber\\
=&\sqrt{1-v^2}\left(f_{0000} + v f_{1100} + v^2 f_{2200} + \dots\right)\label{boosted_example}
\end{align}
and find a truly infinite sum on the right hand side. This is consistent with formula (65) of 
\cite{Baskal:2016rkv}. As a curiosity we note that (\ref{boosted_example}) can be further simplified using the Mehler's formula \cite{zbMATH02750067}. We rewrite 
\begin{equation}
\frac{f_{0000}(\Lambda^{-1}x)}{f_{0}(y)f_0(z)} = \sqrt{1-v^2}\sum_{n=0}^\infty v^n f_{n}(x)f_{n}(t)
\end{equation}
and use the Mehler's formula on the right hand side
\begin{equation}
\sqrt{1-v^2} \sum_{n=0}^\infty v^n f_{n}(x)f_{n}(t) = \pi^{-\frac12}\exp\left(-\frac{1-v}{1+v}\frac{(x+t)^2}{4} -\frac{1+v}{1-v}\frac{(x-t)^2}{4}\right)\,.
\end{equation}

We emphasize again that our method defines a homomorphism from any Lorentz transformation matrix $\Lambda$ to the representation $D_{\multiN}^{\multiM} (\Lambda)$. In case of a boost parameterized by $\vec{v} = (v_x, v_y, v_z)$, we can either directly use (\ref{DmatrixGenD}), or see that the integral (\ref{master_integral}) becomes
\begin{align}
I = \pi^2&\sqrt{1-\vec{v}^2}\nonumber
\exp \Bigg[2 \Big( p^{{0}}q^{{0}} \sqrt{1-\vec{v}^2} - p^{{0}}p^{{1}} v_x - p^{{0}}p^{{2}}v_y - p^{{0}}p^{{3}}v_z + q^{{0}}q^{{1}} v_x + q^{{0}}q^{{2}} v_y + q^{{0}}q^{{3}} v_z\nonumber\\
+&p^{{1}}q^{{1}}\frac{v_y^2 + v_z^2 + v_x^2\sqrt{1-\vec{v}^2}}{\vec{v}^2} - p^{{1}} q^{{2}} \frac{v_xv_y}{\vec{v}^2}(1-\sqrt{1-\vec{v}^2}) - p^{{1}} q^{{3}} \frac{v_xv_z}{\vec{v}^2}(1-\sqrt{1-\vec{v}^2})\nonumber\\
-& p^{{2}} q^{{1}} \frac{v_xv_y}{\vec{v}^2}(1-\sqrt{1-\vec{v}^2}) + p^{{2}}q^{{2}}\frac{v_x^2 + v_z^2 + v_y^2\sqrt{1-\vec{v}^2}}{\vec{v}^2}- p^{{2}} q^{{3}} \frac{v_yv_z}{\vec{v}^2}(1-\sqrt{1-\vec{v}^2})\nonumber\\
-& p^{{3}} q^{{1}} \frac{v_xv_z}{\vec{v}^2}(1-\sqrt{1-\vec{v}^2}) -  p^{{3}} q^{{2}} \frac{v_yv_z}{\vec{v}^2}(1-\sqrt{1-\vec{v}^2}) + p^{{3}}q^{{3}}\frac{v_x^2 + v_y^2 + v_z^2\sqrt{1-\vec{v}^2}}{\vec{v}^2} \Big)\Bigg]\,,
\end{align}
from which it is possible to extract the representation matrix $D_{\multiN}^{\multiM} (\vec{v})$ through the same procedure as above.

\subsection{Representation matrices for rotations in $D=4$}
In the case of pure rotations, we use $\Lambda_{\mu}{}_\nu = R_\mu{}_\nu$, with $R_{00} = -1, R_{0i}=R_{i0} = 0, R_{ij}\neq 0$, where $R_{ij}$ is an orthogonal rotation matrix.
The integral (\ref{master_integral}) becomes very simple;
\begin{equation}\label{master_integral_rotations}
I = \pi^{D/2} 
\exp\left[2p_0q_0\right]
\exp\left[2p_i q_j R_{ji}\right]\,.
\end{equation}
Through the same procedure, we can write down a general representation matrix for an arbitrary $(D-1)$-dimensional spatial rotation in $D$ dimensions.
\begin{align}\label{DmatrixRotation3+1b}
D_{\multiN}^{\multiM}(R) = &\,
\delta_{-n_0+\sum_{i} n_{i}}^{-m_0+\sum_{i} m_{i}}
\delta^{m_0}_{n_0}
\sqrt{\prod_{i} n_{i}! m_{i}!} 
{\sum_{\left\lbrace n_{ij} \right\rbrace}}'
\prod_{ij} \frac{ R_{ij}^{n_{ij}} }{n_{ij}!}\,.
\end{align}
Similarly to above, by
${\sum'_{\left\lbrace n_{ij} \right\rbrace}}$
we denote $(D-1)\times (D-1)$ sums over each $n_{ij}$ from 0 to $\infty$, where the prime $'$ on the sum denotes that the values of $n_{ij}$ are restricted to those that satisfy the constraints:
\begin{align}\label{DmatrixDc}
\sum_j n_{ij} &= m_i \nonumber\\
\sum_i n_{ij} &= n_j 
\end{align}
The sums and products involving $i$ and $j$ range from $1$ to $D-1$. 

In case we restrict our attention solely to $SO(D)$, i.e. to $D$-dimensional rotations in $D$-dimensional Euclidean space, the formulas for representation matrices are of the same form (with the same form of constraints (\ref{DmatrixDc})) except they do not contain the factor $\delta^{m_0}_{n_0}$ and sums and products involving $i$ and $j$ range from $1$ to $D$: 
\begin{align}\label{DmatrixRotation3+0b}
D_{\multiN}^{\multiM}(R) = &\,
\delta_{\sum_{i} n_{i}}^{\sum_{i} m_{i}}
\sqrt{\prod_{i} n_{i}! m_{i}!} 
{\sum_{\left\lbrace n_{ij} \right\rbrace}}'
\prod_{ij} \frac{ R_{ij}^{n_{ij}} }{n_{ij}!} 
\end{align}

Returning to the $SO(1,3)$ case, for a generic rotation $R_{\mu\nu}$ the matrices $D(R)$ can be written as
\begin{align}\label{DmatrixRotation3+1c}
D_{\multiN}^{\multiM}(R) = &\,
\delta_{-n_0+n_1+n_2+n_3}^{-m_0+m_1+m_2+m_3}
\delta^{m_0}_{n_0}
\sqrt{n_1! n_2! n_3! m_1! m_2! m_3!} \times \nonumber\\
& \times
{\sum_{n_{11}=0}^{n_1}}
{\sum_{n_{12}=0}^{n_1}}
{\sum_{n_{21}=0}^{n_2}}
{\sum_{n_{22}=0}^{n_2}}
\frac{ R_{11}^{n_{11}} }{n_{11}!} 
\frac{ R_{12}^{n_{12}} }{n_{12}!} 
\frac{ R_{21}^{n_{21}} }{n_{21}!} 
\frac{ R_{22}^{n_{22}} }{n_{22}!} \times\nonumber \\
& \times
\frac{ R_{13}^{m_1 - n_{11} - n_{12}} }{ \left(m_1 - n_{11} - n_{12} \right)!} 
\frac{ R_{23}^{m_2 - n_{21} - n_{22}} }{ \left(m_2 - n_{21} - n_{22} \right)!} \times\nonumber \\
& \times
\frac{ R_{31}^{n_1 - n_{11} - n_{21}} }{ \left(n_1 - n_{11} - n_{21} \right)!} 
\frac{ R_{32}^{n_2 - n_{12} - n_{22}} }{ \left(n_2 - n_{12} - n_{22} \right)!}\times\nonumber \\
& \times
\frac{ R_{33}^{n_3 - m_1 - m_2 + n_{11} + n_{12} + n_{21} + n_{22}} }
{ \left( n_3 - m_1 - m_2 + n_{11} + n_{12} + n_{21} + n_{22} \right)!}\,.
\end{align}
For clarity and further uses, we provide the representation matrices for rotations  of angle $\theta$ around the $x,y$ and $z$ axes separately:\footnote{Here, we used the following notation and conventions for rotations. E.g.\ for rotations around the $z$ direction:
		\begin{align}
		D_{n_0n_1n_2n_3}^{m_0m_1m_3m_3}(\theta\hat{z}) \equiv D_{n_0n_1n_2n_3}^{m_0m_1m_3m_3}(\Lambda(\theta\hat{z})) \equiv D_{n_0n_1n_2n_3}^{m_0m_1m_3m_3}(R(\theta\hat{z}))
		\end{align}
		with
		\begin{align}
		R(\theta\hat{z}){}^\mu{}{}_\nu \equiv 
		\begin{pmatrix}
		1 & 0 & 0 & 0 \\
		0 & \cos\theta & -\sin\theta & 0 \\
		0 & \sin\theta & \cos\theta & 0 \\
		0 & 0 & 0 & 1
		\end{pmatrix}.
		\end{align}}
\begin{align}\label{Drotx}
&	D_{n_0n_1n_2n_3}^{m_0m_1m_2m_3}(\theta\hat x)=
\delta_{-n_0+n_1+n_2+n_3}^{-m_0+m_1+m_2+m_3}\delta_{n_0}^{m_0}\delta_{n_1}^{m_1}\nonumber \sqrt{n_2!n_3!m_2!m_3!}\times
\\&\qquad\qquad\qquad
\times	\sum_{k=0}^{n_2}\frac{(-1)^{m_2-k}}{k!(n_2-k)!}\frac{(\cos\theta)^{2k + m_3 - n_2}(\sin\theta)^{n_2+m_2-2k}}{(m_3-n_2+k)!(m_2-k)!}
\end{align}
\begin{align}\label{Droty}
&	D_{n_0n_1n_2n_3}^{m_0m_1m_2m_3}(\theta\hat y)=
\delta_{-n_0+n_1+n_2+n_3}^{-m_0+m_1+m_2+m_3}\delta_{n_0}^{m_0}\delta_{n_2}^{m_2}\nonumber \sqrt{n_1!n_3!m_1!m_3!}\times
\\&\qquad\qquad\qquad
\times	\sum_{k=0}^{n_3}\frac{(-1)^{m_3-k}}{k!(n_3-k)!}\frac{(\cos\theta)^{2k + m_1 - n_3}(\sin\theta)^{n_3+m_3-2k}}{(m_1-n_3+k)!(m_3-k)!}
\end{align}
\begin{align}\label{rotation_z_4d}
&	D_{n_0n_1n_2n_3}^{m_0m_1m_2m_3}(\theta\hat z)=
\delta_{-n_0+n_1+n_2+n_3}^{-m_0+m_1+m_2+m_3}\delta_{n_0}^{m_0}\delta_{n_3}^{m_3}\nonumber \sqrt{n_1!n_2!m_1!m_2!}\times
\\&\qquad\qquad\qquad
\times	\sum_{k=0}^{n_1}\frac{(-1)^{m_1-k}}{k!(n_1-k)!}\frac{(\cos\theta)^{2k + m_2 - n_1}(\sin\theta)^{n_1+m_1-2k}}{(m_2-n_1+k)!(m_1-k)!}\,.
\end{align}
Note that one can easily constrain the representation matrix (\ref{rotation_z_4d}) to cover only rotations of Hermite functions on a Euclidean plane by setting $m_0=n_0=m_3=n_3=0$, and thus obtain a representation of $SO(2)$.

In formulas written here factorials of a negative number can appear in denominators (e.g.\ in (\ref{rotation_z_4d}) when $m_1<n_1$ for $k=n_1$). Such terms are excluded from the sums, which is consistent with the definition of the factorial in terms of the gamma function; an infinity in the denominator renders such terms effectively zero. 

The matrices (\ref{Drotx})-(\ref{rotation_z_4d}) are also infinite dimensional, but for spatial rotations, due to the global factor of $\delta_{n_0}^{m_0}$, we find a non-negative invariant number $n=n_1+n_2+n_3$. This makes it obvious that the rotation matrices can be reduced to sectors labeled by $n$, which are finite dimensional themselves, as there is only a finite number of ways to sum $n_1,n_2,n_3$ into a non-negative number.
\section{Lorentz Lie algebra in $D=4$}\label{sec:lie4d}
To find the generators for the Lie algebra in the representation furnished by Hermite functions, we use the convention 
\begin{equation}
D = \exp(K\psi)
\end{equation}
which gives, symbolically
\begin{equation}
K = \frac{\partial D}{\partial \psi}\Big|_{\psi=0}\,.
\end{equation}
Since the matrices $D$ were orthogonal by construction, the generators will be antisymmetric
\begin{equation}
K^T = - K\,.
\end{equation}
For any one-parameter subgroup of boosts in a particular direction of the Lorentz Lie group, rapidity $\psi = \tanh^{-1}(v)$ is the canonical coordinate \cite{choquet1982analysis}.

In the 4-dimensional case, the three generators of boosts are
\begin{align}\label{boost_generators_upper_indices}
K_1{}_{n_0n_1n_2n_3}^{m_0m_1m_2m_3} =& \delta_{-n_0 + n_1+n_2+n_3}^{-m_0+m_1+m_2+m_3}\delta_{n_2}^{m_2}\delta_{n_3}^{m_3}\left(\delta^{m_1}_{n_1+1}\sqrt{(n_0+1)(n_1+1)} - \delta^{m_1}_{n_1-1}\sqrt{n_1n_0}\right)\nonumber\\
K_2{}_{n_0n_1n_2n_3}^{m_0m_1m_2m_3} =& \delta_{-n_0 + n_1+n_2+n_3}^{-m_0+m_1+m_2+m_3}\delta_{n_1}^{m_1}\delta_{n_3}^{m_3}\left(\delta^{m_2}_{n_2+1}\sqrt{(n_0+1)(n_2+1)} - \delta^{m_2}_{n_2-1}\sqrt{n_2n_0}\right)\nonumber\\
K_3{}_{n_0n_1n_2n_3}^{m_0m_1m_2m_3} =& \delta_{-n_0 + n_1+n_2+n_3}^{-m_0+m_1+m_2+m_3}\delta_{n_1}^{m_1}\delta_{n_2}^{m_2}\left(\delta^{m_3}_{n_3+1}\sqrt{(n_0+1)(n_3+1)} - \delta^{m_3}_{n_3-1}\sqrt{n_3n_0}\right)\,,
\end{align}
and the three generators of rotations are
\begin{align}
J_1{}^{m_0m_1m_2m_3}_{n_0n_1n_2n_3} =& \delta^{-m_0+m_1+m_2+m_3}_{-n_0+n_1+n_2+n_3}\delta^{m_1}_{n_1}\delta^{m_0}_{n_0}\left(\delta^{m_2}_{n_2-1}\sqrt{n_2(n_3+1)}-\delta^{m_2}_{n_2+1}\sqrt{(n_2+1)n_3}\right)\nonumber\\
J_2{}^{m_0m_1m_2m_3}_{n_0n_1n_2n_3} =& \delta^{-m_0+m_1+m_2+m_3}_{-n_0+n_1+n_2+n_3}\delta^{m_2}_{n_2}\delta^{m_0}_{n_0}\left(\delta^{m_3}_{n_3-1}\sqrt{n_3(n_1+1)}-\delta^{m_3}_{n_3+1}\sqrt{(n_3+1)n_1}\right)\nonumber\\
J_3{}^{m_0m_1m_2m_3}_{n_0n_1n_2n_3} =& \delta^{-m_0+m_1+m_2+m_3}_{-n_0+n_1+n_2+n_3}\delta^{m_3}_{n_3}\delta^{m_0}_{n_0}\left(\delta^{m_1}_{n_1-1}\sqrt{n_1(n_2+1)}-\delta^{m_1}_{n_1+1}\sqrt{(n_1+1)n_2}\right)\label{rotation_generators_upper_indices}\,.
\end{align}
The Lie algebra satisfies the expected products\footnote{A more often used convention in physics for the exponential map from the Lie algebra to the Lie group elements is $D = e^{-i\theta J}$, which gives the familiar products \begin{equation}
	\begin{aligned}
	{\left[J_{i}, J_{j}\right] } &=i \epsilon_{i j k} J_{k} \\
	{\left[J_{i}, K_{j}\right] } &=i \epsilon_{i j k} K_{k} \\
	{\left[K_{i}, K_{j}\right] } &=-i \epsilon_{i j k} J_{k}\,.
	\end{aligned}
	\end{equation}}
\begin{equation}
\begin{aligned}
{\left[J_{i}, J_{j}\right] } &= \epsilon_{i j k} J_{k} \\
{\left[J_{i}, K_{j}\right] } &= \epsilon_{i j k} K_{k} \\
{\left[K_{i}, K_{j}\right] } &=- \epsilon_{i j k} J_{k}\,.
\end{aligned}
\end{equation}
This representation of the Lorentz Lie algebra is not irreducible, and it is a non-trivial problem to reduce it completely. We already noted one level of reducibility, which comes from fixing the number $N=-n_0+n_1+n_2+n_3$. We leave open the question of reducing the representation further.

For future reference, we note the values of the Casimir operators of the Lorentz group $SO(1,3)$ and the rotation group $SO(3)$ in this representation.
There are two Casimir elements of the Lorentz group, both are quadratic
\begin{align}
c_1 = \vec{J}^{\,2} - \vec{K}^{\,2},\quad c_2 = \vec{J}\cdot\vec{K}
\end{align}
In the representation defined above, they become\footnote{Note that each term in the sum contains a product of five deltas, and that one of the deltas is redundant. E.g.\ the overall delta $\delta^{-m_0+m_1+m_2+m_3}_{-n_0+n_1+n_2+n_3}$ can be deduced from the Kronecker delta's in each term in the bracket. We keep such deltas to make the invariance of $-m_0+m_1+m_2+m_3$ manifest.}
\begin{align*}
c_1{}^{m_0m_1m_2m_3}_{n_0n_1n_2n_3} =&\delta^{-m_0+m_1+m_2+m_3}_{-n_0+n_1+n_2+n_3}\times\Big(\\
-2&\delta^{m_0}_{n_0}\delta^{m_1}_{n_1}\delta^{m_2}_{n_2}\delta^{m_3}_{n_3}\,a_{n_0n_1n_2n_3}\\ 
+&\delta^{m_0}_{n_0}\delta^{m_1}_{n_1}\delta^{m_2}_{n_2-2}\delta^{m_3}_{n_3+2}\sqrt{(n_2-1)n_2(n_3+1)(n_3+2)}\\ 
+&\delta^{m_0}_{n_0}\delta^{m_1}_{n_1}\delta^{m_2}_{n_2+2}\delta^{m_3}_{n_3-2}\sqrt{(n_2+1)(n_2+2)(n_3-1)n_3}\\
+&\delta^{m_0}_{n_0}\delta^{m_1}_{n_1+2}\delta^{m_2}_{n_2}\delta^{m_3}_{n_3-2}\sqrt{(n_3-1)n_3(n_1+1)(n_1+2)}\\ 
+&\delta^{m_0}_{n_0}\delta^{m_1}_{n_1-2}\delta^{m_2}_{n_2}\delta^{m_3}_{n_3+2}\sqrt{(n_3+1)(n_3+2)(n_1-1)n_1}\\
+&\delta^{m_0}_{n_0}\delta^{m_1}_{n_1-2}\delta^{m_2}_{n_2+2}\delta^{m_3}_{n_3}\sqrt{(n_1-1)n_1(n_2+1)(n_2+2)}\\ 
+&\delta^{m_0}_{n_0}\delta^{m_1}_{n_1+2}\delta^{m_2}_{n_2-2}\delta^{m_3}_{n_3}\sqrt{(n_1+1)(n_1+2)(n_2-1)n_2}\\
+&\delta^{m_0}_{n_0-2}\delta^{m_1}_{n_1-2}\delta^{m_2}_{n_2}\delta^{m_3}_{n_3}\sqrt{(n_0-1)n_0(n_1-1)n_1} \\
+&\delta^{m_0}_{n_0+2}\delta^{m_1}_{n_1+2}\delta^{m_2}_{n_2}\delta^{m_3}_{n_3}\sqrt{(n_0+1)(n_0+2)(n_1+1)(n_1+2)}\\
+&\delta^{m_0}_{n_0-2}\delta^{m_1}_{n_1}\delta^{m_2}_{n_2-2}\delta^{m_3}_{n_3}\sqrt{(n_0-1)n_0(n_2-1)n_2} \\
+&\delta^{m_0}_{n_0+2}\delta^{m_1}_{n_1}\delta^{m_2}_{n_2+2}\delta^{m_3}_{n_3}\sqrt{(n_0+1)(n_0+2)(n_2+1)(n_2+2)}\\
+&\delta^{m_0}_{n_0-2}\delta^{m_1}_{n_1}\delta^{m_2}_{n_2}\delta^{m_3}_{n_3-2}\sqrt{(n_0-1)n_0(n_3-1)n_3} \\
+&\delta^{m_0}_{n_0+2}\delta^{m_1}_{n_1}\delta^{m_2}_{n_2}\delta^{m_3}_{n_3+2}\sqrt{(n_0+1)(n_0+2)(n_3+1)(n_3+2)}\,\Big)\numberthis
\end{align*}
with 
\begin{equation}
a_{n_0n_1n_2n_3} = -3 (2 + n_1 + n_2 + n_3) - 
2 (n_2 n_3 + n_1 n_2 + n_1 n_3 + n_0 (3 + n_1 + n_2 + n_3))\,.
\end{equation}
The second Casimir operator vanishes
\begin{align*}
c_2{}^{m_0m_1m_2m_3}_{n_0n_1n_2n_3} =&\delta^{-m_0+m_1+m_2+m_3}_{-n_0+n_1+n_2+n_3}\times\\
\Big(-&\delta^{m_0}_{n_0-1}\delta^{m_1}_{n_1-1}\delta^{m_2}_{n_2+1}\delta^{m_3}_{n_3-1}\sqrt{n_0n_1(n_2+1)n_3}\\
+&\delta^{m_0}_{n_0+1}\delta^{m_1}_{n_1+1}\delta^{m_2}_{n_2+1}\delta^{m_3}_{n_3-1}\sqrt{(n_0+1)(n_1+1)(n_2+1)n_3}\\
+&\delta^{m_0}_{n_0-1}\delta^{m_1}_{n_1-1}\delta^{m_2}_{n_2-1}\delta^{m_3}_{n_3+1}\sqrt{n_0n_1n_2(n_3+1)}\\
-&\delta^{m_0}_{n_0+1}\delta^{m_1}_{n_1+1}\delta^{m_2}_{n_2-1}\delta^{m_3}_{n_3+1}\sqrt{(n_0+1)(n_1+1)n_2(n_3+1)}\\
-&\delta^{m_0}_{n_0-1}\delta^{m_2}_{n_2-1}\delta^{m_3}_{n_3+1}\delta^{m_1}_{n_1-1}\sqrt{n_0n_2(n_3+1)n_1}\\
+&\delta^{m_0}_{n_0+1}\delta^{m_2}_{n_2+1}\delta^{m_3}_{n_3+1}\delta^{m_1}_{n_1-1}\sqrt{(n_0+1)(n_2+1)(n_3+1)n_1}\\
+&\delta^{m_0}_{n_0-1}\delta^{m_2}_{n_2-1}\delta^{m_3}_{n_3-1}\delta^{m_1}_{n_1+1}\sqrt{n_0n_2n_3(n_1+1)}\\
-&\delta^{m_0}_{n_0+1}\delta^{m_2}_{n_2+1}\delta^{m_3}_{n_3-1}\delta^{m_1}_{n_1+1}\sqrt{(n_0+1)(n_2+1)n_3(n_1+1)}\\
-&\delta^{m_0}_{n_0-1}\delta^{m_3}_{n_3-1}\delta^{m_1}_{n_1+1}\delta^{m_2}_{n_2-1}\sqrt{n_0n_3(n_1+1)n_2}\\
+&\delta^{m_0}_{n_0+1}\delta^{m_3}_{n_3+1}\delta^{m_1}_{n_1+1}\delta^{m_2}_{n_2-1}\sqrt{(n_0+1)(n_3+1)(n_1+1)n_2}\\
+&\delta^{m_0}_{n_0-1}\delta^{m_3}_{n_3-1}\delta^{m_1}_{n_1-1}\delta^{m_2}_{n_2+1}\sqrt{n_0n_3n_1(n_2+1)}\\
-&\delta^{m_0}_{n_0+1}\delta^{m_3}_{n_3+1}\delta^{m_1}_{n_1-1}\delta^{m_2}_{n_2+1}\sqrt{(n_0+1)(n_3+1)n_1(n_2+1)}\,\Big)
\\=&\,0\numberthis\,.
\end{align*}

The group $SO(3)$ has a single Casimir element $\vec{J}^{\,2}$, which is given by
\begin{align*}
(\vec{J})^2\,{}^{m_1m_2m_3}_{n_1n_2n_3} =& \delta^{m_1+m_2+m_3}_{n_1+n_2+n_3}\times\\
\Big(-2&\delta^{m_1}_{n_1}\delta^{m_2}_{n_2}\delta^{m_3}_{n_3}(n_1+n_2+n_3 + n_1n_2 + n_2n_3 + n_3n_1)\\ 
+&\delta^{m_1}_{n_1}\delta^{m_2}_{n_2-2}\delta^{m_3}_{n_3+2}\sqrt{(n_2-1)n_2(n_3+1)(n_3+2)}\\ 
+&\delta^{m_1}_{n_1}\delta^{m_2}_{n_2+2}\delta^{m_3}_{n_3-2}\sqrt{(n_2+1)(n_2+2)(n_3-1)n_3}\\
+&\delta^{m_1}_{n_1+2}\delta^{m_2}_{n_2}\delta^{m_3}_{n_3-2}\sqrt{(n_3-1)n_3(n_1+1)(n_1+2)}\\ 
+&\delta^{m_1}_{n_1-2}\delta^{m_2}_{n_2}\delta^{m_3}_{n_3+2}\sqrt{(n_3+1)(n_3+2)(n_1-1)n_1}\\
+&\delta^{m_1}_{n_1-2}\delta^{m_2}_{n_2+2}\delta^{m_3}_{n_3}\sqrt{(n_1-1)n_1(n_2+1)(n_2+2)}\\ 
+&\delta^{m_1}_{n_1+2}\delta^{m_2}_{n_2-2}\delta^{m_3}_{n_3}\sqrt{(n_1+1)(n_1+2)(n_2-1)n_2}\,\Big)\,.\label{so(3)Casimir}\numberthis
\end{align*}

\subsection{Diagonalization of $J_3$}\label{sec:diagJ3}
In this last part, we focus on a specific application of the newfound generators. In the basis (\ref{multi_dimensional_hermite}), none of the generators above are diagonal. The physical motivation for building this representation came from an expansion we proposed in \cite{Cvitan:2021qvm} for the MHS field, and we aim to study the helicity content of the MHS fields through such an expansion. 

Since the conventional way of building the little group for a massless field is by choosing the little momentum to be in the $z$ direction we explicitly perform the diagonalization of $J_3$. It turns out that a similar problem is encountered in a treatment of the two dimensional isotropic quantum harmonic oscillator (see e.g.\ \cite{NikiforovUvarovSuslov}) and it also can be applied in the context of digital image analysis for precise image rotation \cite{Park2008, ParkAccurate2009, Reynolds2018}. The further intent of this section is therefore to apply formulas obtained in previous sections to a known example, which serves as a check and also to connect the results of this work with the existing literature. Details of the calculations are delegated to the appendices, and here we present the results.

An element of $L^2(\mathbb{R}^{4})$
\begin{equation}
\Phi = \sum_{\multiM=0}^\infty p^{m_0m_1m_2m_3}f_{m_0m_1m_2m_3}
\end{equation}
is an eigenvector of the rotation operator $J_3 \cdot \Phi = \lambda \Phi\,$ if the components $p^{m_0m_1m_2m_3}$ satisfy the equation
\begin{equation}
J_3{}^{m_0m_1m_2m_3}_{n_0n_1n_2n_3}p^{n_0n_1n_2n_3} = \lambda p^{m_0m_1m_2m_3}\,,
\end{equation}
where the repeated indices are summer over. Since the chosen rotation generator leaves invariant indices $n_0,n_3$, we suppress them and focus only on the part of interest
\begin{equation}\label{2d_generator_rotations}
J_3{}^{m_1m_2}_{n_1n_2} = \delta^{m_1+m_2}_{n_1+n_2}\left(\delta^{m_1}_{n_1-1}\sqrt{n_1(n_2+1)}-\delta^{m_1}_{n_1+1}\sqrt{(n_1+1)n_2}\right)\,.
\end{equation}
The eigenvalue equation reduces to $J_3{}^{m_1m_2}_{n_1n_2} C^{n_1n_2} = \lambda C^{m_1m_2}$, or in terms of the new components $C^{m_1m_2}$
\begin{align}\label{recursioncmm}
C^{m_1+1,m_2-1}\sqrt{(m_1+1)m_2} -C^{m_1-1,m_2+1}\sqrt{m_1(m_2+1)} = \lambda C^{m_1,m_2}\,.
\end{align}
Due to the presence of $\delta^{m_1+m_2}_{n_1+n_2}$ in (\ref{2d_generator_rotations}), the number $r=n_1+n_2$ is invariant under the action of the rotation generator, which makes any sector of the vector space with a fixed $r$ finite dimensional. This enables numerical calculations of the coefficients and the eigenvalues, and we report some of them as an explicit example, in the form $C^{m_1m_2}_{(r,\lambda)}$ where $r=m_1+m_2$ and $\lambda$ is the eigenvalue.
\begin{align}
C^{m_1m_2}_{(0,0)} =& \delta^{m_1}_{0}\delta^{m_2}_{0} \label{C_example_first}\\
C^{m_1m_2}_{(1,-i)} =& \frac{1}{\sqrt{2}}(\delta^{m_1}_{0}\delta^{m_2}_1 -i \delta^{m_1}_{1}\delta^{m_2}_0)\\
C^{m_1m_2}_{(1,i)} =& \frac{1}{\sqrt{2}}(-i \delta^{m_1}_{0}\delta^{m_2}_1 + \delta^{m_1}_{1}\delta^{m_2}_0)\label{C_example_last}	
\end{align}
To express the solution in a closed form we define
\begin{align}
C_{r,n_1}^{m_1}(\beta) 
&\equiv  
(-i)^{ n_1+m_1} \left( \sin\frac{\beta}{2} \right)^{ n_1+m_1} \left( \cos\frac{\beta}{2} \right)^{r-( n_1+m_1)}
\sqrt{\binom{r}{m_1} \binom{r}{n_1}} \,\,
\times\nonumber
\\
&\qquad\qquad\times
{}_2F{}_1(-n_1,-m_1,-r;\sin^{-2} \frac{\beta}{2})\label{Cbetadef}
\\\nonumber
&=
i^{n_1-m_1} d^{r/2}{}_{n_1-r/2,m_1-r/2}(\beta)  
\\\nonumber
&=
i^{n_1-m_1} K_{n_1}^{(\beta)}(m_1,r) 
\end{align}
where the second and third equalities are a consequence of the definitions of the Wigner $d$-functions and the functions $K_{n_1}^{(\beta)}(m_1,r)$ defined and reviewed in Appendix~\ref{funreview}. Since $K_{n_1}^{(\beta)}(m_1,r)$ are the ``normalized'' Kravchuk polynomials, it is easy to check that
$C_{r,n_1}^{m_1}(\beta)$ inherit the orthogonality, completeness and symmetry:
\begin{align}
& \sum_{m_1 = 0}^r C_{r,n_1}^{m_1}{}^*(\beta) C_{r,n_1'}^{m_1}(\beta) = \delta_{n_1n_1'}  \label{Cbort}\\
& \sum_{n_1 = 0}^r C_{r,n_1}^{m_1}{}^*(\beta) C_{r,n_1}^{m_1'}(\beta) = \delta_{m_1m_1'}	\label{Cbcompl} \\
&	\qquad C_{r,n_1}^{m_1}(\beta) = C_{r,m_1}^{n_1}(\beta) \label{Cbsymm}\,.
\end{align}
We show in Appendix~\ref{J3eigenvectors} that the argument $\beta=\pi/2$ provides for the eigenvectors of $J_3$ with the eigenvalue $\lambda =  i(2k-r)$ for each $r$ and $k$:
\begin{align}
C_{(r,\lambda)}^{m_1, m_2} 
= \delta^{m_1+m_2}_{r} C_{r,k}^{m_1}\left( \frac\pi2 \right)\, , 
\quad \lambda =  i(2k-r), 
\quad 0 \leq k \leq r
\end{align}
When $\beta=\pi/2$ we use the index $k$ instead of $n_1$ to emphasize this (i.e.\ $J_3{}_{k}^{k'} = \sum_{m_1',n_1'} 
	C_{r,k}^{n_1'} \left( \frac\pi2 \right)
	J_3{}^{m_1'}_{n_1'} 
	C_{r,k'}^{m_1'}{}^*\left( \frac\pi2 \right)$ is diagonal).

The finite rotation matrix for states of fixed $r=m_1+m_2$ 
is found as follows:
\begin{align}
\exp( \beta J_3 )^{m_1}_{n_1} &= 
\sum_{k,k',m_1',n_1'} 
C_{r,k}^{n_1}{}^*\left( \frac\pi2 \right) C_{r,k}^{n_1'} \left( \frac\pi2 \right)
\exp( \beta J_3 )^{m_1'}_{n_1'} 
C_{r,k'}^{m_1'}{}^*\left( \frac\pi2 \right) C_{r,k'}^{m_1}\left( \frac\pi2 \right)  \nonumber\\
&=
\sum_{k} 
C_{r,k}^{n_1}{}^* \left( \frac\pi2 \right)
\exp( i \beta (2k-r) )
C_{r,k}^{m_1}\left( \frac\pi2 \right)  \nonumber\\
&=	i^{n_1-m_1} C_{r,n_1}^{m_1}( 2\beta ) =	d^{r/2}{}_{n_1-r/2,m_1-r/2}( -2\beta ) \label{expJ3}
\end{align}
where we use the completeness relation to isolate $\exp( \beta J_3 )_{k}^{k'} = \exp( i \beta (2k-r) )\delta_{k}^{k'}$ and we use the formula (\ref{Cfin2}) in the second line. This agrees with \cite{Park2008,ParkAccurate2009,Reynolds2018}\footnote{
	The factor of $2$ that multiplies $\beta$ on the right hand side of (\ref{expJ3}) is consistent with the fact that the rotation of an element of the Hermite basis by $2\pi$ must be an identity, i.e.\ it is consistent since $d^{1/2}(2\pi)=-1$ and $d^{1/2}(4\pi)=1$.}. 

For an arbitrary value of $r$ we obtain the following result (i.e.\ the exponentiation of (\ref{2d_generator_rotations})) 
\begin{align}\label{expJ3r}
\exp( \beta J_3 )^{m_1m_2}_{n_1n_2}
&= 
\delta^{m_1+m_2}_{n_1+n_2} 
d^{\frac12(m_1+m_2)}{}_{\frac12(n_1-n_2),\frac12(m_1-m_2)}( -2\beta )\,, 
\end{align}
which also agrees with (\ref{rotation_z_4d})
\begin{align}\label{expJ3r1}
D_{n_0n_1n_2n_3}^{m_0m_1m_2m_3}(\beta \hat z)
&= 
\delta^{m_0}_{n_0} \delta^{m_1+m_2}_{n_1+n_2} 
\delta^{m_3}_{n_3}
d^{\frac12(m_1+m_2)}{}_{\frac12(n_1-n_2),\frac12(m_1-m_2)}( -2\beta )\,.
\end{align}

\section{Conclusion}
We have constructed the infinite dimensional unitary representation of the Lorentz group for multi-dimensional Hermite functions that form the basis of the representation space, and thus provided for the first time an explicit expression for the transformation matrix for an arbitrary element of the Lorentz group in such a basis, in any number of dimensions. 

Even though our original motivation was rooted in the study of MHS fields, 
the results obtained are, however, self-standing and could be used without reference to the MHS formalism. 
In particular, our results complete those obtained in \cite{Ruiz:1974uf,Rotbart:1981xf}, and therefore might furnish a tool in studies of the relativistic harmonic oscillator in arbitrary dimensions; the matrices (\ref{DmatrixGenD}) provide an overlap between wave functions of the relativistic oscillator in the laboratory frame and any other Lorentz frame. 

In section~\ref{sec:lie4d}, we have calculated the generators of the Lorentz Lie algebra in this representation and calculated the Casimir elements of the Lorentz group in $D=4$ and the rotation group in $D=3$. 
In the subsection \ref{sec:diagJ3}, for a special case of a two dimensional rotation, we provide a neat derivation for a finite rotation of the Hermite basis. The derivation makes use of a related basis where the rotation generator $J_3$ is diagonal. Such formulas are used in digital graphics to achieve accurate rotations of two dimensional images by means of a  Hermite basis expansion, as in \cite{ParkAccurate2009,Reynolds2018}. The derivation here employs Kravchuk polynomials and makes a connection to the particular hypergeometric function representation. Takeaway message of this subsection (and appendices) is the practicality of these quantities since they are regular for the parameters of interest, as well as the convenience of using the Wigner $d$-functions to express the result for the rotation formula as opposed to the generalized associated Legendre functions.

In the context of field theory and the MHS formalism, 
we see possible applications in formulating infinite component fields, akin to \cite{Majorana:1932chs, Casalbuoni:2006fa, Abers:1967zza}. Finally,
the obtained Lie algebra generators (\ref{boost_generators_upper_indices})-(\ref{rotation_generators_upper_indices}) will be beneficial in explicitly constructing the representation of the quartic Casimir operator of the Poincar\'e group for an on-shell MHS field using the little group approach. We leave this analysis to our future work.

\section*{Acknowledgments}
The research of P.D.P.\ has been supported by the University of Rijeka under the project uniri-prirod-18-256 and uniri-iskusni-prirod-23-222. The research of M.P.\ has been supported by the University of Rijeka under the project uniri-mladi-prirod-23-43 3159. The research of S.G.\ has been supported by a BIRD-2021 project (PRD-2021) and by the  PRIN Project n.~2022ABPBEY, ``Understanding quantum field theory through its deformations''. The work in this paper draws upon research initially presented in the doctoral dissertation of M.P.\ \cite{Paulisic:2023lam} and extends beyond the thesis with further investigations and expansions.
\appendix

\section{Kravchuk and Jacobi polynomials and generalized associated Legendre functions}
\label{funreview}
Here we would like to review the relation of the Wigner $d$-functions to Kravchuk and Jacobi polynomials as well as to the generalized associated Legendre functions. Kravchuk polynomials are an example of orthogonal polynomials and have been treated extensively in mathematical texts \cite{NikiforovUvarovSuslov}, \cite{Koekoek2010}, \cite{VilenkinKlimyk}. They are closely related to the perhaps more widely known Jacobi polynomials (see below). All these functions can be represented in terms of the hypergeometric functions in various ways (some of which are defined only for a subset of parameter space needed in physical applications). The one we use in (\ref{Cbetadef}) and below, often used to define the Kravchuk polynomials, is particularly useful for representing the Wigner $d$-functions since it is nonsingular for the parameters of interest\footnote{Comparing the representation (\ref{Khyp}) with the 
following one (often used to express the Jacobi polynomials): $k_n^{(p)}(x,N)  = P_n^{(x-n,N-x-n)}(1-2p) = ( (x+1-n)_n / n! ) \,\, {}_2F{}_1(-n,N+1-n,x+1-n;p)$ we see that the former avoids singularities that appear on the right hand side of the latter. E.g.\ for $N=n=1$, $x=0$, the former contains the finite ${}_2F{}_1(-1,0,-1;1/p)$, whereas the latter contains the singular ${}_2F{}_1(-1,1,0;p)$. In that sense, (\ref{Khyp}) is more practical for our purposes.}.

When relating the Wigner $d$-functions to other functions mentioned here it is important to note that the Wigner $d$-functions are periodic, with the period of $4\pi$. Kravchuk and Jacobi polynomials, of variable $p$, where $p=\sin^2\frac{\beta}{2}$, have a period of $2\pi$ in $\beta$. When normalized, in a particular way (see footnote \ref{footnotesqrtp}), they are known as Kravchuk functions (which has a period of $4\pi$ in $\beta$) and are proportional to the Wigner $d$-functions. In the case of generalized associated Legendre functions (that are defined as functions of $z=\cos \beta$) the rotation angle on one hand can be complex but on another hand is restricted to $0\le \real \beta < \pi$.

The definition of the Kravchuk polynomials is
\begin{equation}\label{Kbin}
k_n^{(p)}(x,N) 
=  \sum_{i=0}^{n} (-1)^{n-i} \binom{N-x}{n-i} \binom{x}{i} p^{n-i} (1-p)^{i}\,, 
\end{equation}
and they are related to the hypergeometric functions as\footnote{Some authors (\cite{VilenkinKlimyk}, \cite{Koekoek2010}) define the Kravchuk polynomials as 
	${}_2F{}_1(-n,-x,-N;1/p)$.}%
\begin{align}\label{Khyp}
k_n^{(p)}(x,N) 
&=  (-p)^n \binom{N}{n} \,\, {}_2F{}_1(-n,-x,-N;1/p) \,.
\end{align}
This definition matches the definition of the Jacobi polynomials
\begin{align}\label{Phyp}
	P_n^{(a,b)}(z) 
	=  \sum_{s} \binom{n+a}{s} \binom{n+b}{n-s} \left( \frac{z-1}{2} \right)^{n-i} \left( \frac{z+1}{2} \right)^{i}
\end{align}
if we use
\begin{align}
p = \frac{1-z}{2}
\,,\quad
x = n + a
\,,\quad
N = 2n + a + b\,\,.
\end{align}
In this way we get \cite{NikiforovUvarovSuslov}
\begin{align}
k_n^{(p)}(x,N) = P_n^{(x-n,N-x-n)}(1-2p) \,.
\end{align}
To compare these polynomials with the Wigner $d$-functions we express them in terms of $\beta$ which is related to $p$ (or $z$) as follows:
\begin{align}
z = \cos\beta\,\,,\quad
p = \frac{1-z}{2} = \sin^2\frac{\beta}{2}\,\,,\quad
1-p = \frac{1+z}{2} = \cos^2\frac{\beta}{2}\,.
\end{align}
Next, we ``normalize'' the Kravchuk polynomials, i.e.\ multiply them by $c_n^{(\beta)}(x,N)$ consisting of the square root of their weight divided by the square root of their norm (listed in \cite{NikiforovUvarovSuslov})%
\footnote{\label{footnotesqrtp}In this step one needs to choose the branches of the square roots.
		An obvious choice, the one used in (\ref{constc}) consists in taking $\sqrt{p}=\sin \frac{\beta}{2}$ and $\sqrt{1-p}=\cos \frac{\beta}{2}$. This makes $K_n^{(\beta)}(x,N)$ proportional to the Wigner $d$-functions as in (\ref{dK}) for all (real) $\beta$.}.
The resulting function (sometimes called the Kravchuk function, see e.g.\  \cite{Atakishiyev:97}) is
\begin{align}
K_n^{(\beta)}(x,N) 
&= c_n^{(\beta)}(x,N)\;  	k_n^{(\sin^2 \frac{\beta}{2})}(x,N)  \\
&= c_n^{(\beta)}(x,N) \;	P_n^{(x-n,N-x-n)}(\cos\beta) \label{KJac} \,.
\end{align}
where
\begin{align}
c_n^{(\beta)}(x,N) \label{constc}
&= \binom{N}{x}^{1/2} \binom{N}{n}^{-1/2} 
\left( \sin \frac{\beta}{2} \right)^{x-n}
\left( \cos \frac{\beta}{2} \right)^{N-n-x}
\end{align}
We note that $c_n^{(\beta)}(x,N)$ has a period of $4\pi$ in $\beta$.
The Wigner $d$-functions are defined as
\begin{align}
d^l{}{}_{m',m}(\beta) & = 
\sqrt{(l+m')!(l-m')!(l+m)!(l-m)!} \times \\\nonumber
& \qquad\qquad \times 
\sum_s \frac{
	(-1)^{m'-m+s}
	\left( \sin \frac{\beta}{2} \right) ^{2s+m'-m}
	\left( \cos \frac{\beta}{2} \right) ^{2l-2s-m'+m}
}{s!(l+m-s)!(l-m'-s)!(s+m'-m)!}
\end{align}
and it is well known that the relation to the Jacobi polynomials reads
\begin{align} \label{dJac}
d^l{}{}_{m',m}(\beta) & = 
\sqrt{\frac{(l+m)!(l-m)!}{(l+m')!(l-m')!}} 
\left( \sin \frac{\beta}{2} \right) ^{m-m'}
\left( \cos \frac{\beta}{2} \right) ^{m'+m}
\times \\\nonumber
& \qquad\qquad \times 
P_{l-m}^{(m-m',m+m')}(\cos\beta)
\end{align}
Comparing (\ref{constc}) and (\ref{KJac}) with (\ref{dJac}) one gets (c.f.\ \cite{NikiforovUvarovSuslov}) the relations between the Wigner $d$-functions and the Kravchuk function valid for all (real) $\beta$: 
\begin{align}\label{dK}
&d^l{}{}_{n,m}(\beta) = (-)^{m-n}d^l{}{}_{m,n}(\beta) 
= K_{l-m}^{(\beta)}(l-n,2l)
\\\nonumber
&= K_{l+n}^{(\beta)}(l+m,2l)
= (-)^{m-n}K_{l-n}^{(\beta)}(l-m,2l)
= (-)^{m-n}K_{l+m}^{(\beta)}(l+n,2l)
\end{align}
We show in the Appendix~\ref{J3eigenvectors} how Kravchuk matrices naturally appear when constructing eigenvectors and eigenvalues of the rotation operator for the $2$-dimensional Hermite basis, and use that in (\ref{expJ3}) to calculate the finite rotation.
Here, we also mention the generalized associated Legendre functions (see e.g.\ Chapter~6 of \cite{VilenkinKlimyk}). These are related to the Jacobi polynomials in the following way
\begin{align}\label{GALegChirJac}
& P{}^l_{mn}(\cos \beta) = \sqrt{\frac{(l-m)!(l+m)!}{(l-n)!(l+n)!}}  
\left( \sin \frac{\beta}{2} \right) ^{m-n}
\left( \cos \frac{\beta}{2} \right) ^{m+n}
\times \\\nonumber
& \qquad\qquad \times 
P_{l-m}^{(m-n,m+n)}(\cos\beta)
\end{align}
Here, the right hand side is of the same form as (\ref{dJac}), but due to the $\cos \beta$ on the left hand side the definition (\ref{GALegChirJac}) is valid only for $0\le \real \beta < \pi$. 
In this way, since $z=\cos\beta$, $P{}^l_{mn}$ viewed as a function of $z$ is single valued on the complex plane $z$.
So comparing (\ref{constc}) and (\ref{KJac}) with (\ref{GALegChirJac}), we see that generalized associated Legendre functions match the Wigner $d$-functions only for $0 \leq \beta < \pi$: 
\begin{align}
& P{}^l_{mn}(\cos \beta) = K_{l-m}^{(\beta)}(l-n,2l) = d^l{}{}_{n,m}(\beta)
, \qquad 0 \leq \beta < \pi
\end{align}
For that reason we find it more convenient to use $d^l{}{}_{n,m}(\beta)$ instead of $P{}^l_{mn}(\cos \beta)$ to express the results such as (\ref{expJ3r}).
\section{$J_3$ eigenvectors and Kravchuk polynomials}
\label{J3eigenvectors}
To solve (\ref{recursioncmm}) we use the following redefinition for the coefficients $C^{m_1m_2}_{(r,\lambda)}$ which will enable us to rewrite it in a simpler way. 
\begin{equation}\label{prefactor}
C_{r,k}^{m_1 m_2} = N_{r,k}  \frac{e^{i \pi (m_2-m_1)/4} }{\sqrt{(2m_1)!!(2m_2)!!}}P_{r,k}^{m_1,m_2}\,
\end{equation}
where $r = m_1 + m_2$, while $k$ will correspond to the eigenvalue $\lambda$, with the exact dependency to be determined.
The normalization constant $N_{r,k}$ does not affect (\ref{recursioncmm}). 
Furthermore, we choose to work with a fixed $r$, and drop the $m_2$ labels since we can express $m_2=r-m_1$. 
\begin{equation}\label{prefactor2}
C_{r,k}^{m_1} = N_{r,k}  \frac{e^{i \pi (r-2 m_1)/4} }{\sqrt{(2m_1)!!(2(r-m_1))!!}}P_{r,k}^{m_1}
\end{equation}
The eigenvalue equation becomes
\begin{equation}
\label{recursion2}
m_1 P_{r,k}^{m_1-1}+(r-m_1) P_{r,k}^{m_1+1} -  i \lambda  P_{r,k}^{m_1} = 0
\end{equation}
Surprisingly, the solution to this equation is given by the Kravchuk matrices \cite{krawtchouk1929generalisation,krawtchouk1933distribution}. They are defined as
\begin{equation}
K^{(N)}_{ij} = \sum_{k=0}^N (-1)^k\binom{j}{k}\binom{N-j}{i-k}
\end{equation}
and we can use them in the following way
\begin{equation}
\label{kravchuk_1st_use}
P_{r,k}^{m_1} = 
K^{(r)}_{k\,m_1} =  \sum_{i=0}^{r} (-1)^{i} \binom{m_1}{i} \binom{r-m_1}{k-i}\,.
\end{equation}
To prove that this is a solution and to find the eigenvalues, we will first rewrite the Kravchuk matrices in terms of the Kravchuk polynomials. Since there is a connection of the Kravchuk polynomials to the hypergeometric function, we will be able to re-express (\ref{kravchuk_1st_use}) using the hypergeometric function. Then, we will use the known formulas for the hypergeometric function and prove our solution.

Using (\ref{Khyp}) in (\ref{kravchuk_1st_use}) we set $x \rightarrow m_1$, $N \rightarrow m_1+m_2=r$, $N-x \rightarrow m_2$, $n \rightarrow k$ and $p\to 1/2$, from which it follows
\begin{align}
k_k^{(1/2)}(m_1,r) =&\sum_{i=0}^{k} (-1)^{k-i}\binom{m_1}{i} \binom{r-m_1}{k-i}  2^{-k} \\
=&(-1)^{k} 2^{-k} K^{(r)}_{k\,m_1}\\
=& (-1)^{k} 2^{-k} \binom{r}{k} \,\, {}_2F{}_1(-k,-m_1,-r;2) \label{relation_kF}\,.
\end{align}
Therefore
\begin{align}
P_{r,k}^{m_1} &= \binom{r}{k}\,\, {}_2F{}_1(-k,-m_1,-r;2) \\
&= (-1)^{k} 2^{k} k_k^{(1/2)}(m_1,r)
\,.
\end{align}
It is most easily seen from the equation above that the integer parameter $k$, due to the binomial coefficient, can range from $0$ to $r$.
We now use the consecutive recurrence relation \cite{functions.wolfram.com} for the hypergeometric functions :
\begin{align}
(b - c)\, {}_2F{}_1(a,b-1,c;z) + (c - 2 b + (b - a) z)\, {}_2F{}_1(a,b,c;z)&\nonumber \\= b (z - 1) \,{}_2F{}_1(a,b+1,c;z)&\,.
\end{align}
To adapt this recurrence relation to our problem, we set
\begin{equation}
z=2,\quad a=-k,\quad b=-m_1,\quad c=-r=-m_1-m_2\,.
\end{equation} The recurrence relation becomes
\begin{align}
m_2\,\, {}_2F{}_1(-k,-(m_1+1),-r;2) - (r - 2 k)\,\, {}_2F{}_1(-k,-m_1,-r;2)\nonumber& \\= -m_1 \,\, {}_2F{}_1(-k,-(m_1-1),-r;2)&\,.
\end{align}
This is now identical to the equation (\ref{recursion2}) with the eigenvalue $\lambda = i(2k - r)$.
As we have established above, the parameter $k$ ranges from $0$ to $r$, which means that the eigenvalue $\lambda$ can, for a certain choice of $r$, attain values
\begin{equation}
\lambda_{r} = -ir,-i(r+1),...,i(r-1), ir\,.
\end{equation}
Finally, we choose the normalization factors $N_{r,k}$ of $C_{r,k}^{m_1}$  such that:
\begin{align}\label{diagonal_rotation2}
C_{r,k}^{m_1} 
&=  i^{ k-m_1} 
\frac{\sqrt{\frac{1}{2^r}\binom{r}{m_1}}}{\frac{1}{2^{k}}\sqrt{\binom{r}{k}} }  k_k^{(1/2)}(m_1,r) 
\,.
\end{align}
Besides solving the diagonalization problem (\ref{recursioncmm}), such $C_{r,k}^{m_1}$ satisfy the following properties of orthogonality, completeness and symmetry
\begin{align}
& \sum_{m_1 = 0}^r C_{r,k}^{m_1}{}^* C_{r,k'}^{m_1} = \delta_{kk'}  \label{Cort}\\
& \sum_{k = 0}^r C_{r,k}^{m_1}{}^* C_{r,k}^{m_1'} = \delta_{m_1m_1'}	\label{Ccompl} \\
&	\qquad C_{r,k}^{m_1} = C_{r,m_1}^{k} \label{Csymm}
\end{align}
These properties are inherited from the corresponding properties of the Kravchuk polynomials, as follows.
The symmetry of the Kravchuk polynomials reads
\begin{align}
(-2)^{k} \binom{r}{m_1} k_k^{(1/2)}(m_1,r)  = 	(-2)^{m_1}  \binom{r}{k} k_{m_1}^{(1/2)}(k,r)
\end{align}
It follows from the identity ${}_2F{}_1(-k,-m_1,-r;2) = {}_2F{}_1(-m_1,-k,-r;2)$ applied to (\ref{relation_kF}), and 
one can check that it implies (\ref{Csymm}). This property is manifest if we express 
(\ref{diagonal_rotation2}) in terms of the function ${}_2F{}_1$
\begin{align}\label{diagonal_rotation3}
C_{r,k}^{m_1} 
&=       
(-i)^{k+m_1} 2^{-r/2}
\sqrt{\binom{r}{m_1} \binom{r}{k}} \,\, {}_2F{}_1(-k,-m_1,-r;2)
\,.
\end{align}
The orthogonality of the Kravchuk polynomials reads (see e.g.\  \cite{Atakishiyev:97})
\begin{align}
\sum_{m_1 = 0}^r \frac{1}{2^{r}} \binom{r}{m_1} k_k^{(1/2)}(m_1,r) k_{k'}^{(1/2)}(m_1,r) = \frac{1}{2^{2k}} \binom{r}{k} \delta_{kk'}
\end{align}
and can be readily used to check (\ref{Cort}).
Finally, (\ref{Ccompl}) is a consequence of (\ref{Cort}) and (\ref{Csymm}).

We now note that $C_{r,m_1'}^{m_1}$ are a special case for $\beta=\frac{\pi}{2}$ of a more general function $C_{r,m_1'}^{m_1}( \beta )$ defined in (\ref{Cbetadef}). As shown in the second line of (\ref{Cbetadef}), $C_{r,m_1'}^{m_1}$ is proportional to the Wigner $d$-functions.
Using one of the properties of $d^{l}{}_{mn}(\beta)$ known as Wigner's trick (see e.g.\ \cite{Aubert2013})
\begin{align}
i^{s_1'-s_1} \sum_{s=-l}^l	
d^{l}{}_{s_1',s}\left(  \frac\pi{2}  \right)
\exp(i \beta s) 
d^{l}{}_{s,s_1}\left(  -\frac\pi{2}  \right) &= d^{l}{}_{s_1',s_1}( \beta ) \label{dfin2}
\end{align}
and using (\ref{Cbetadef}) we obtain the following identity relating   $C_{r,m_1'}^{m_1}\left(  \frac\pi{2}  \right)$  and $C_{r,m_1'}^{m_1}(\beta)$
\begin{align}
\sum_{k=0}^r	
C_{r,k}^{m_1'}{}^*\left(  \frac\pi{2}  \right) \exp(i \frac{\beta}{2} (2k-r)) C_{r,k}^{m_1}\left(  \frac\pi{2}  \right) &=
i^{m_1'-m_1} C_{r,m_1'}^{m_1}( \beta ) \label{Cfin2}
\end{align}
\vfill
\bibliography{references}
\end{document}